\documentclass[12pt]{article}

\usepackage{amsmath,amsfonts,epsfig,mathrsfs,todonotes,yfonts}
\usepackage{eufrak}
\usepackage{tikz}
\usetikzlibrary{shapes,arrows}

\usepackage{xcolor}
\usepackage{cite}
\usepackage{stmaryrd}
\usepackage{mdframed}
\usepackage[top=2.5cm, bottom=2.5cm, left=2.75cm, right=2.75cm]{geometry} 
\usepackage{dsfont}
\numberwithin{equation}{section}
\usepackage{hyperref}
 \hypersetup{
     colorlinks=true,
     linkcolor=black,
     filecolor=blue,
     citecolor=blue,      
     urlcolor=cyan,
     }
\usepackage[most]{tcolorbox}

\newcommand{\re}[1] {(\ref{#1})}

\newcommand{\pa}{\partial} 

\newcommand{\ber}{\begin{eqnarray}}
\newcommand{\eer}[1]{\label{#1}\end{eqnarray}}
\newcommand{\eero}{\end{eqnarray}}

\newcommand{\balg}{\begin{align}}
\newcommand{\ealg}{\end{align}}

\newcommand{\beq}{\begin{equation}}
\newcommand{\eeq}{\end{equation}}
\newcommand{\bea}{\begin{eqnarray}}
\newcommand{\eea}{\end{eqnarray}}

\newcommand{\nn}{\nonumber}
\newcommand{\na}{\nabla}

\def\HollowBox #1#2{{\dimen0=#1 \advance\dimen0 by -#2
       \dimen1=#1 \advance\dimen1 by #2
        \vrule height #1 depth #2 width #2
        \vrule height 0pt depth #2 width #1
        \llap{\vrule height #1 depth -\dimen0 width \dimen1} 
       \hskip -#2
       \vrule height #1 depth #2 width #2}}

\usepackage{fancyhdr} 

\usepackage{chngcntr}
\counterwithout{equation}{section}
\newcommand{\auth}{\large Ulf Lindstr\"om ${}^{a,b}$\footnote{email: ulf.lindstrom@physics.uu.se}
and {\"O}zg{\"u}r Sar{\i}o\u{g}lu ${}^a$\footnote{email: sarioglu@metu.edu.tr}}
\begin{document}
\rhead{DRAFT}
\begin{flushright}
{\small UUITP-52/22}\\
\vskip 1.5 cm
\end{flushright}

\begin{center}
{\Large{\bf A comment on Metric vs Metric-Affine Gravity}}
\vspace{.75cm}

\auth
\end{center}
\vspace{.5cm}
\vspace{.5cm}
\centerline{${}^a${\it \small Department of Physics, Faculty of Arts and Sciences,}}
\centerline{{\it \small Middle East Technical University, 06800, Ankara, Turkey}}
\vspace{.5cm}
\centerline{${}^b${\it \small Department of Physics and Astronomy, Theoretical Physics, Uppsala University}}
\centerline{{\it \small SE-751 20 Uppsala, Sweden }}

\vspace{1cm}


\centerline{{\bf Abstract}}
\bigskip

\noindent
We consider the sum of the Einstein-Hilbert action and a Pontryagin density (PD) in arbitrary 
even dimension $D\geq 4$. All curvatures are functions of independent affine (torsionless) 
connections only. In arbitrary even dimension, not only in $D=4n$, these first order PD terms are shown 
to be covariant divergences of ``Chern-Simons'' currents. The field equation for the connection 
leads to it being Levi-Civita, and to the metric and affine field equations being 
equivalent to the second order metric theory. This result is a counterexample to the theorem 
stating that purely metric and metric-affine models can only be equivalent for Lovelock theories. 
\vskip .5cm

\vspace{0.5cm}
\small

\renewcommand{\thefootnote}{\arabic{footnote}}
\setcounter{footnote}{0}

\pagebreak
\setcounter{page}{2}

The action we consider in $D=2n$ dimensions reads
\ber
S = \int d^Dx \left( \sqrt{|g|} g^{ab} R_{ab}(\Gamma) +{\textstyle\frac 1 n}  \theta\epsilon^{a_1a_2 \dots a_{D}} 
R^{i_1}{}_{i_2 a_1 a_2} R^{i_2}{}_{i_3 a_3 a_4} \dots R^{i_n}{}_{i_1 a_{D-1} a_D}(\Gamma) \right) \,.
\eer{act1}
In the metric formulation the connection is the Levi-Civita connection and the second term 
vanishes for odd $n$ due to the additional symmetries of the curvature tensor.
The topological nature of the second term becomes manifest if written as $\pa_a K^a$ 
where,\footnote{Here we assume that $\theta$ is constant. In the Chern-Simons modification of 
General Relativity (GR) \cite{Jackiw:2003pm}, it is taken to be a scalar field, much in the spirit of the 
Fradkin-Tseytlin term in string theory. This leads to interesting deviations from GR. It follows from 
the results in the present paper that promoting $\theta$ to a scalar field will give modifications of 
GR also for odd $n$, as in, e.g., six dimensions. Doing so will alter the solution for the connection, 
which will now depend on $\theta$.} schematically,
\ber
K^{a_1} \approx\theta \epsilon^{a_1 a_2 \dots a_{D}} \Gamma^{i_1}{}_{a_2 i_2} \Big( 
R^{i_2}{}_{i_3 a_3 a_4} \dots R^{i_n}{}_{i_1 a_{D-1} a_D} + \Gamma^2 R^{n-2} 
+ \dots + \Gamma^{D-2} \Big)~.
\eer{}
The $D=4$ expression reads \cite{Jackiw:2003pm,Ertem:2012bhz}
\ber
K^{a} =  \theta \epsilon^{abcd} \Gamma^{i}{}_{bj} \Big(R^{j}{}_{i c d} 
- \frac{2}{3} \Gamma^{j}{}_{c k} \Gamma^{k}{}_{d i} \Big)~,
\eer{}
while in $D=6$ we have 
\ber\nn
  K^{a} =\frac{2}{3} \theta \epsilon^{abcdef}\Gamma^{i}_{~bj}\Big(R^j_{~kcd}R^k_{~ief}- R^j_{~kcd}\Gamma^{k}_{~er}\Gamma^{r}_{~fi}+\frac 2 5\Gamma^{j}_{~ck}\Gamma^{k}_{~dp}\Gamma^{p}_{~eq}\Gamma^{q}_{~fi}\Big)~.
  \eer{}

The field equations that follow from varying the connection in \re{act1} read 
\ber
-2\na_{a} \Big (\theta \epsilon^{a b a_3 \dots a_{D}} R^{d}{}_{c a_3 a_4} \dots 
R^{i_n}{}_{i_1 a_{D-1} a_D} \Big) + \na_{c} (\sqrt{|g|} g^{d b}) 
- \delta^d{}_c \na_{a} (\sqrt{|g|} g^{ab} ) = 0 \,.
\eer{f2}
The first term vanishes due to Bianchi identities and the remaining terms may be massaged to give
\ber
\na_{c} (\sqrt{|g|} g^{d b}) = 0 \,,
\eer{f2}
which implies that the connection is the Levi-Civita connection. This means that the action 
\re{act1} is completely equivalent to its purely metric form where the connection is Levi-Civita 
from the outset.

In \cite{Exirifard:2007da} it is argued that the only case when the metric and metric-affine 
formulations of gravity are equivalent is Lovelock gravity \cite{Lovelock:1971yv}. However, 
gravity amended with a Pontryagin term as described above is not a Lovelock gravity and thus 
constitutes a counterexample. It must be noted though that the discussion in 
\cite{Exirifard:2007da} is carried out at the level of the equations of motion, whereas the statement 
in this note is at the level of the action.

\vspace {.5cm}

\noindent
{\em Note added}: While we were writing these results up, a paper,\cite{Sulantay:2022sag}, appeared on the net which contains the metric-affine description of the $4D$ version of \re{act1}. Similar results also appeared in \cite{Boudet:2022wmb}.\\

\noindent{\bf Acknowledgments}\\
The research of U.L. is supported in part by the 2236 Co-Funded 
Scheme2 (CoCirculation2) of T\"UB{\.I}TAK (Project No:120C067)\footnote{\tiny However 
the entire responsibility for the publication is ours. The financial support received from 
T\"UB{\.I}TAK does not mean that the content of the publication is approved in a scientific 
sense by T\"UB{\.I}TAK.}.


\begin{thebibliography}{6666}

\bibitem{Jackiw:2003pm}
R.~Jackiw and S.~Y.~Pi,
``Chern-Simons modification of general relativity,''
\href{https://dx.doi.org/10.1103/PhysRevD.68.104012}{{\em
Phys. Rev. D} {\bfseries 68} (2003), 104012};
\href{http://arxiv.org/abs/gr-qc/0308071}{{\ttfamily arXiv:gr-qc/0308071 [gr-qc]}}.

\bibitem{Ertem:2012bhz}
\"U.~Ertem and \"O.~A\c{c}\i{}k,
``Generalized Chern-Simons Modified Gravity in First-Order Formalism,''
\href{https://dx.doi.org/10.1007/s10714-012-1483-8}{{\em
Gen. Rel. Grav.} {\bfseries 45} (2013), 477-488};
\href{http://arxiv.org/abs/0912.1433}{{\ttfamily arXiv:0912.1433 [gr-qc]}}.

\bibitem{Exirifard:2007da}
Q.~Exirifard and M.~M.~Sheikh-Jabbari,
``Lovelock gravity at the crossroads of Palatini and metric formulations,''
\href{https://dx.doi.org/10.1016/j.physletb.2008.02.012}{{\em 
Phys. Lett. B} {\bfseries 661} (2008), 158-161};
\href{http://arxiv.org/abs/0705.1879}{{\ttfamily arXiv:0705.1879 [hep-th]}}.

\bibitem{Lovelock:1971yv}
D.~Lovelock,
``The Einstein tensor and its generalizations,''\\
\href{https://dx.doi.org/10.1063/1.1665613}{{\em J. Math. Phys.}
{\bfseries 12} (1971), 498-501}.

\bibitem{Sulantay:2022sag}
F.~Sulantay, M.~Lagos and M.~Ba\~nados,
``Chiral Gravitational Waves in Palatini Chern-Simons,''
\href{http://arxiv.org/abs/2211.08925}{{\ttfamily arXiv:2211.08925 [gr-qc]}}.

\bibitem{Boudet:2022wmb}
S.~Boudet, F.~Bombacigno, G.~J.~Olmo and P.~J.~Porfirio,
``Quasinormal modes of Schwarzschild black holes in projective invariant Chern-Simons modified gravity,''
\href{https://dx.doi.org/10.1088/1475-7516/2022/05/032}{{\em JCAP} {\bfseries 05} (2022) no.05, 032};
\href{http://arxiv.org/abs/arXiv:2203.04000}{{\ttfamily arXiv:2203.04000 [gr-qc]}}.

\end{thebibliography}
\end{document}